\documentclass[10pt,conference,twocolumn]{IEEEtran}
\IEEEoverridecommandlockouts

\usepackage{amsmath,amssymb,amsfonts}
\usepackage{algorithmic}
\usepackage{graphicx}
\usepackage{textcomp}
\usepackage{xcolor}
\usepackage[super]{nth}
\usepackage[caption=false, font=footnotesize]{subfig}
\def\BibTeX{{\rm B\kern-.05em{\sc i\kern-.025em b}\kern-.08em
    T\kern-.1667em\lower.7ex\hbox{E}\kern-.125emX}}
\usepackage[backend=biber,style=ieee,doi=false]{biblatex}
\begin{document}

\title{Deep Learning For Experimental Hybrid Terrestrial and Satellite Interference Management\\
\thanks{This work has received funding from the Spanish Ministry of Economy and Competitiveness (Ministerio de Economia	y Competitividad) under project TEC2017-90093-C3-1-R and from the Catalan Government (2017 SGR 1479).}
}

\author{
	\IEEEauthorblockN{Pol Henarejos\IEEEauthorrefmark{1}, Miguel \'{A}ngel V\'{a}zquez\IEEEauthorrefmark{1} and Ana I. P\'{e}rez-Neira\IEEEauthorrefmark{1}\IEEEauthorrefmark{2}}\\
	\IEEEauthorblockA{\IEEEauthorrefmark{1}Centre Tecnol\`ogic de Telecomunicacions de Catalunya (CTTC/CERCA)\\Av. Carl Friedrich Gauss 7, 08860, Castelldefels, Barcelona, Spain\\ Email:\{pol.henarejos,miguel.angel.vazquez,ana.perez\}@cttc.es}
    \IEEEauthorblockA{\IEEEauthorrefmark{2}Signal Theory and Communications Department,\\Universitat Polit\`{e}cnica de Catalunya, Barcelona, Spain}}

\maketitle

\begin{abstract}
Interference Management is a vast topic present in many disciplines. The majority of wireless standards suffer the drawback of interference intrusion and the network efficiency drop due to that. Traditionally, interference management has been addressed by proposing signal processing techniques that minimize their effects locally. However, the fast evolution of future communications makes difficult to adapt to new era. In this paper we propose the use of Deep Learning techniques to present a compact system for interference management. In particular, we describe two subsystems capable to detect the presence of interference, even in high Signal to Interference Ratio (SIR), and interference classification in several radio standards. Finally, we present results based on real signals captured from terrestrial and satellite networks and the conclusions unveil the courageous future of AI and wireless communications.
\end{abstract}


\section{Introduction}
The exponential growth of more demanding needs and more resources has a major impact in the spectrum management. Thanks to recent advances, such as narrower filtering with shaped envelopes \cite{Scaglione1999,Bellanger2008} or cognitive spectrum \cite{Alsarhan2011,Thomas2017}, more services and verticals can coexist smoothly to accommodate new paradigms. 

However, these systems are far to be perfect and may interfere to others. Actually, detecting and managing interference is a paramount task in order to preserve the full efficiency of the operator's network. Detection of interfering signals is a well-studied topic addressed in the last decades \cite{4167652, 317857}. These methods rely on the decision theory of hypothesis testing where a specific knowledge of the signal structure and the channel model is required. Bearing in mind the myriad of current different wireless standards, the development of specific detectors for each signal class becomes a cumbersome task.

In this paper we consider a data driven approach where we aim to perform the signal classification by directly processing the IQ samples. Inspired by the recent applications of deep learning in many fields and, in particular, in signal classification \cite{8357902, 8325299}, this paper proposes the use of Deep Learning for interference classification in satellite systems. Considering an incumbent DVB-S2 satellite signal, we provide an interference classifier able to classify the interference from the well-known cellular standards; namely, LTE, UMTS and GSM.

With the mentioned approach, we avoid the development of different classifiers for each interfering signal which yield to high complex computations and mathematical models. In this context, the suggested procedure might be deployed with low cost equipment in any satellite teleport, satellite user terminal or even on the payload. Numerical results using real data show the potential of the designed deep learning scheme in detecting and classifying the different interfering signals.

To sum up, the contributions of the paper are two. First, we provide a design of an interference detector based on deep neural network coder architecture able to detect an arbitrary signal in presence of a satellite transmission. Second, we introduce the design of deep neural network classifier of cellular transmissions in presence of satellite ones.

\section{System Model}
We aim at detecting and classifying the interferences that are present in satellite communications in different scenarios. We consider the downlink between the satellite and the ground terminal. Without loss of generality, we consider single antenna satellite terminals. In general, we do not perform any assumption on the type of interference. We denote the system model as 
\begin{equation}
y(t)=\sqrt{\rho}\left(x(t)+\sqrt{\gamma^{-1}}i(t)\right)+w(t),
\label{eq:sysmod}
\end{equation}
where $y(t)$ is the received signal, $x(t)$ is the conveyed signal (the intended signal) with unitary power, $i(t)$ is the interference signal from interfering source with unitary power, $w(t)$ is the Additive White Gaussian Noise (AWGN) with unitary power, $\rho$ is the signal to noise ratio (SNR) and $\gamma$ is the signal to interference ratio (SIR).

Before classifying the interference among the considered Radio Access Technologies (RAT), first we need to detect whether the interference is present or not. 

\section{Deep Learning for Interference Detection}
In this case we analyse the probability of detecting an interference. Due to the physical characteristics of broadcast transmissions, it is possible that other signals from other sources interfere with the intended signal. Detecting an interference is not always an easy task, since no information is available neither to the transmitter nor the receiver. Examining the Power Spectrum Density (PSD) may reveal some information on the frequency being interfered. However, this is not always true if the interference’s power is equal or below the power of the intended signal.
In those cases, it is extremely complicated to detect whether an interference exists or not. To circumvent this problem, we propose the use of Machine Learning / Artificial Intelligence (ML/AI) techniques to detect it at the receiver side. In particular, we employ a technique based on Deep Neural Network (DNN) Autoencoder (AE), which uses autoencoders to achieve the objective as we propose in the previous use-case.

Once the input data is similar to the trained data, the autoencoder will also produce similar output data. Thus, the mean squared error (MSE) between the input data and output data provides a metric to decide whether the output is similar or not. The main idea of autoencoding detection is to exploit this feature by training the autoencoder with a signal that does not contain any interference, testing the autoencoder with other signals without interference to obtain useful thresholds and, finally, using it with signals with interference (or not). 

In the presence of different signals (i.e., signals with interference), the MSE produced by the autoencoder is higher compared with the output corresponding to input signals without interferences. In these cases, the autoencoder is not able to produce the same output signal with the same fidelity and, hence, we can exploit it to detect the presence of interference.

\begin{figure}[!ht]
	\centering
	\subfloat[Without interference.]{\includegraphics[width=0.49\linewidth,clip=true]{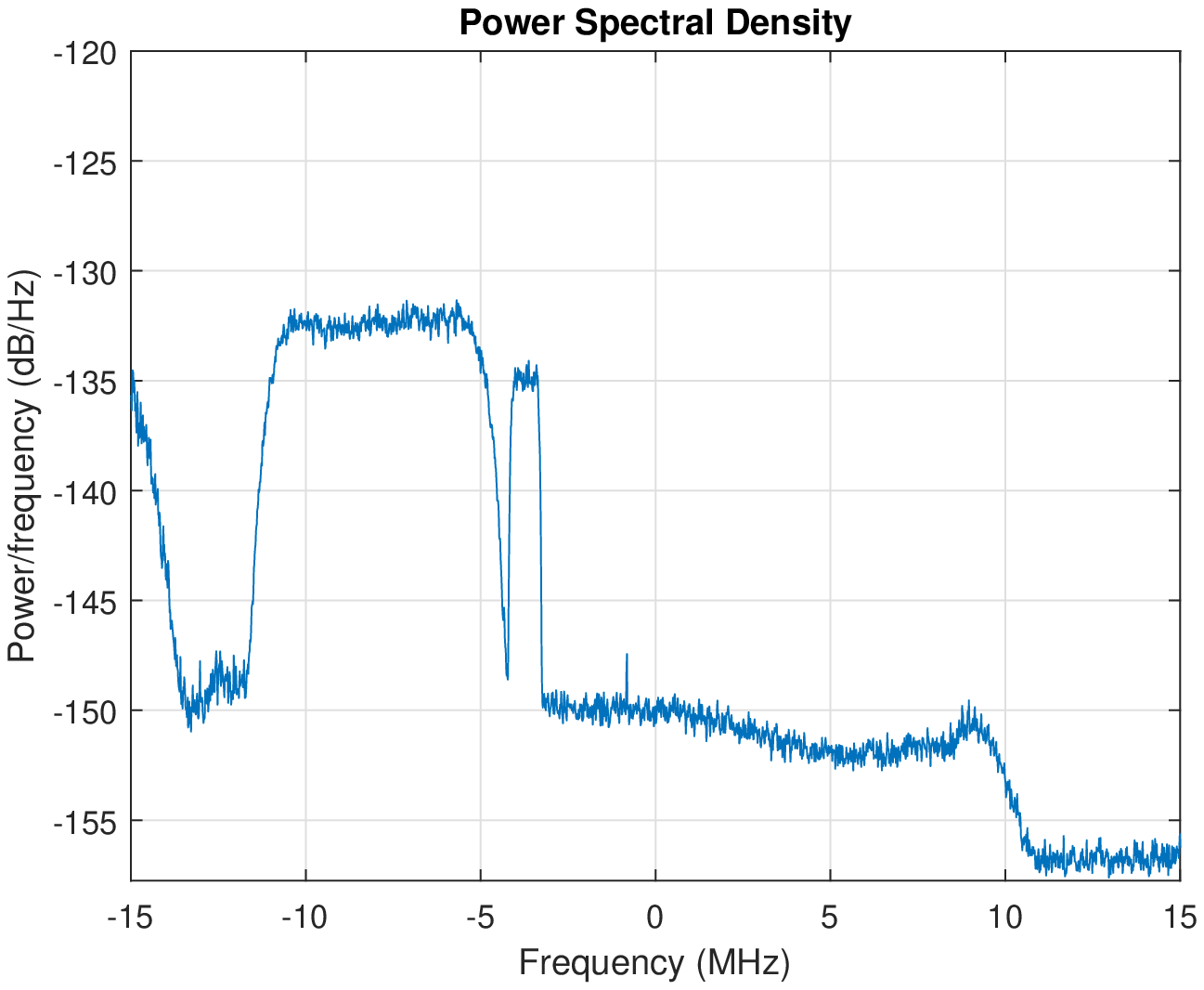}\label{fig:pwelch_noint}}
	\subfloat[With interference, remarked within the circled area.]{\includegraphics[width=0.49\linewidth,clip=true]{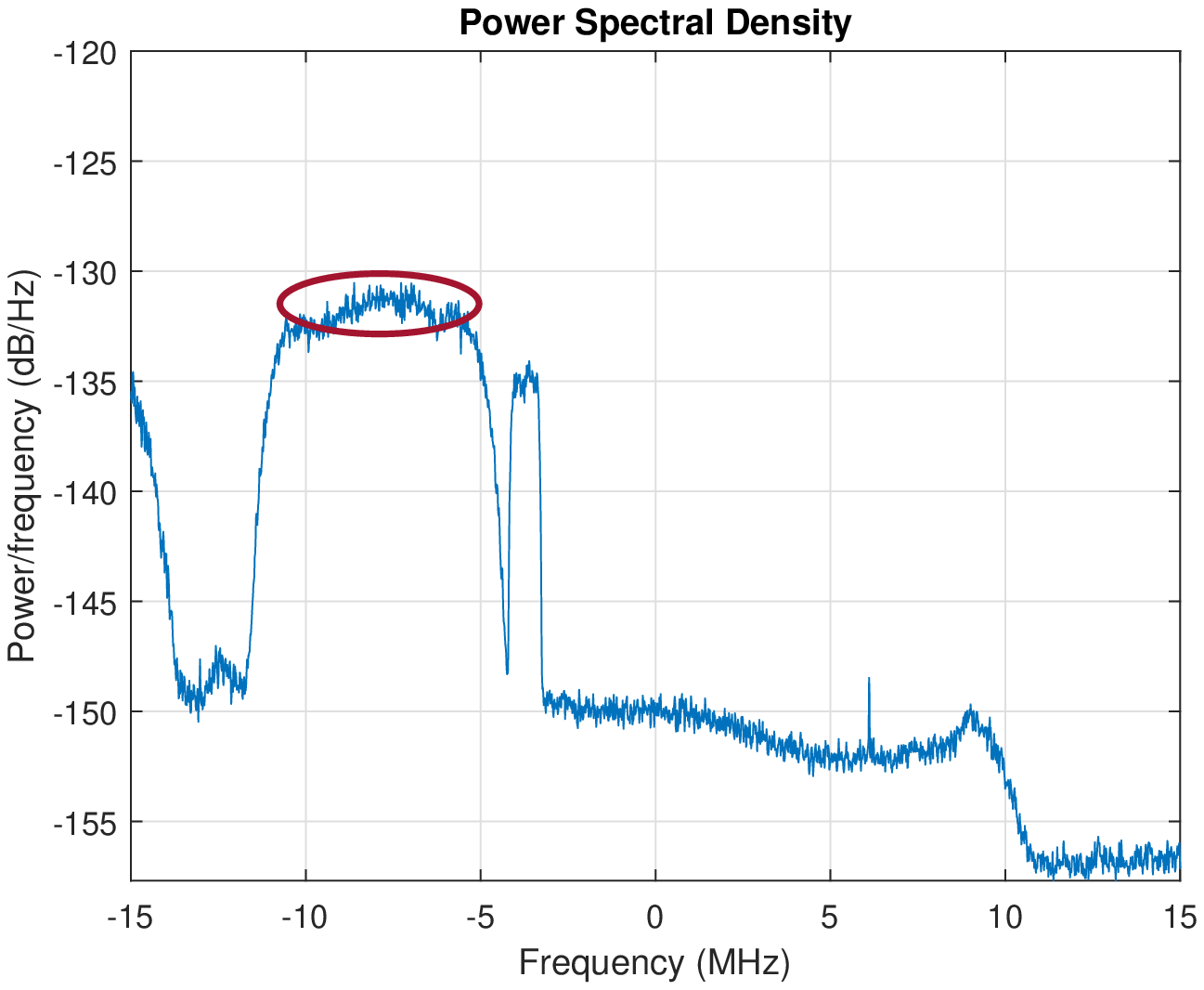}\label{fig:pwelch_yesint}}
	\caption{PSD of the employed sets of real IQ samples.}
	\label{fig:pwelch_noyesint}
\end{figure}

Under these circumstances, we employ two sets of data based on baseband samples (namely, IQ samples) as input data. The first set does not contain any interference and the second set does contain a small interference placed in the centre of the transmitting band.
Fig. \ref{fig:pwelch_noint} depicts the PSD of the set of data without interference. Fig. \ref{fig:pwelch_yesint} depicts the PSD of the set of data with interference. The interference is observable in the middle of the left wide band.

\subsection{Methodology}
First, we divide the first data set in several segments. The first segment is used for the training of the autoencoder, and the rest (test segments) are used to obtain the different averaged thresholds. The rest of segments are used as inputs and we measure the Mean Square Error (MSE) between the input and the produced output. 

The autoencoder is composed by an encoder and a decoder, stacked contiguously at the receiver side. Both functions have the capability to sparse and extract some useful information and, as expected, will depend on the number of hidden layers. Introducing more hidden layers in the autoencoder, will produce an output with less MSE. However, it will increase the probability of false positives since it will be very sensitive to the input data. Hence, for our purpose, we will use a small hidden number of layers. 

Among these aspects, there also exist other parameters that may produce different results. The encoder and decoder transfer functions are with the encoder and decoder, respectively, and are used for avoidance of saturation or smooth the input. Other parameters, such as L2 weight regularizer, sparsity proportion or sparsity regularization coefficient, weight the sparsity of the network. Finally, there is the training function used by the autoencoder for the training. To implement this function, we use the approach described by \cite{Moller1993}, which is referred as scaled conjugate gradient descent function. The regularizer function is used as a cost function to increase the sparsity of the autoencoder. In our case, we use the cost function detailed by \cite{Olshausen1997}.

Obviously, if we use the training segment, the MSE tends to $0$. But it is not the case when we use the rest of segments, since the signals in the inputs are different from the training. For each test segment we measure the MSE and, at the end of the calibration process, we collect all the values and obtain a vector of several MSE. We call it \emph{MSE vector}.

\section{Deep Learning for Interference Classification}
Upon an interference detection, it is also important to classify the interference to narrow the search of possible interfering sources. This task is not always easy since it requires hard dedication to identify the source of interference. Usually, this task is performed manually by operators. The transmission of allowed signals is stopped during this task in order to maximize the chance of identifying the source. 
Obviously, stopping the communication while this task is carried is not the optimal solution. Hence, we propose the implement this task by using a DNN model trained previously with different waveforms corresponding to different Radio Access Networks, such as LTE, UMTS or GSM. 
By training a DNN with a DVB-S2 signal interfered with LTE, UTMS or GSM, we are able to perform classification based on these patterns. For this purpose, we design a Neural Network with a single layer based on Long Short-Term Memory (LSTM) Network. This network is suitable for training networks based on time series signals, such as radio signals\cite{Hochreiter1997}.

\begin{figure}[!ht]
	\centering
	\includegraphics[width=0.9\linewidth,clip=true]{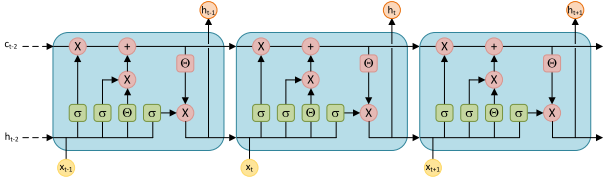}
	\caption{Long Short Term Memory Network schema.}
	\label{fig:LSTM}
\end{figure}

LSTM networks manage to keep contextual information of inputs by integrating a loop that allows information to flow from one step to the next. In contrast to Convolution Neural Networks (CNN), LSTM networks do not use neurons as processing units, but elementary cells. Fig. \ref{fig:LSTM} displays the unrolled loop composing the elementary cell composing these type of neural networks. The input vector is denoted by $x_t$, the output vector is denoted by $h_t$ and the hidden states, $c_t$, are propagated jointly with the output to the next cell. 

In LSTM, there are three gates to control the flow of information: the input gate, the output gate and the forget gate. The first balances the amount of information that goes into the cell; the second, the amount of information that goes out from the cell; the third, the amount of information that remains inside the cell. Usually, the hyperbolic tangent function, $\textrm{tanh}$, is employed to update the cell and hidden states. The sigmoid function is used for the activation of the three gates. Both are denoted by $\Theta$ and $\sigma$, respectively.

\subsection{Methodology}
Under these circumstances, we employ two sets of sampled data. The first set is used for the training and the second is used for the testing. Both sets are composed by real DVB-S2 signal at baseband. Additional to the DVB-S2 signal, which is used as intended signal, we also captured LTE, UMTS and GSM signals at different bands.
 
All captured signals span $50$ MHz of bandwidth and contain radio signals from different operators. The sets are composed by different sample vectors of 512 size that contain only one type interference. For instance, the first vector contains a DVB-S2 signal plus a scaled ($\beta$) LTE signal. The second vector contains the DVB-S2 plus a scaled UMTS signal. The third contains the DVB-S2 plus a scaled GSM signal. And it repeats cyclically. 

The scale parameter is used to obtain a range of different SIRs, denoted by $\gamma$ in \ref{eq:sysmod}. Hence, each set contains the same vector scaled by different values of SIR. It is important to remark that, since all the signals are commercial signals captured in civil environments, these signals contain the typical noise present in all devices. Thus, for the sake of formality, the SIR is expressed as
\begin{equation}
\hat{\gamma}=\frac{P_{\textrm{DVB-S2}}+N_{\textrm{DVB-S2}}}{\beta\left(I_{\textrm{LTE/UMTS/GSM}}+N_{\textrm{LTE/UMTS/GSM}}\right)} 
\end{equation}
where $P_{\textrm{DVB-S2}}$ is the power of the intended signal, $N_{\textrm{DVB-S2}}$ is the noise of the captured DVB-S2 signal, $\beta$ is the scale parameter to obtain different SIRs, $I_{\textrm{LTE/UMTS/GSM}}$ is the power of the interference and $N_{\textrm{LTE/UMTS/GSM}}$ is the noise of the terrestrial captured signal.

\begin{figure}[!ht]
	\centering
	\subfloat[DVB-S2]{\includegraphics[width=0.49\linewidth,clip=true]{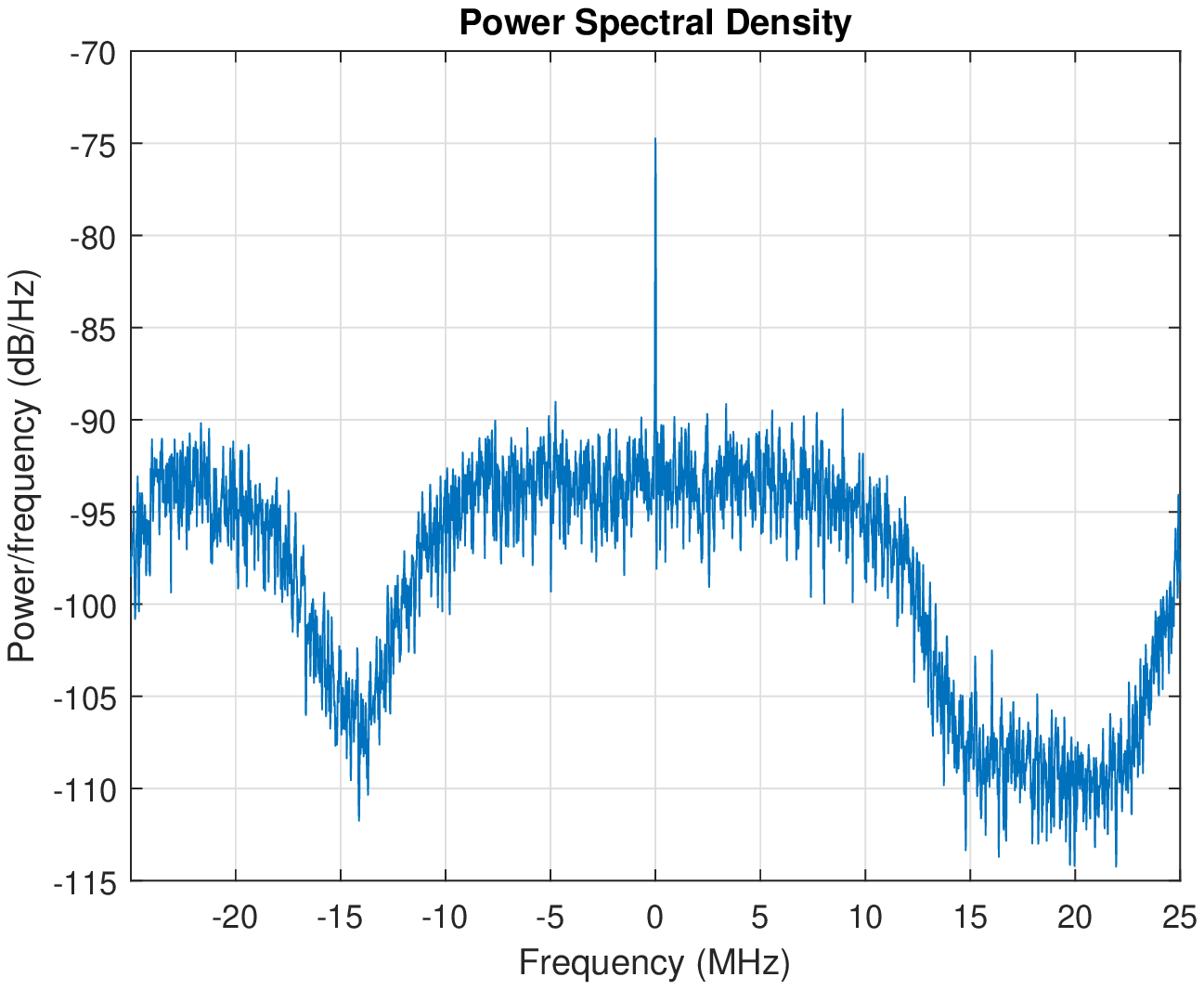}\label{fig:pwelch_dvbs2}}
	\subfloat[LTE]{\includegraphics[width=0.49\linewidth,clip=true]{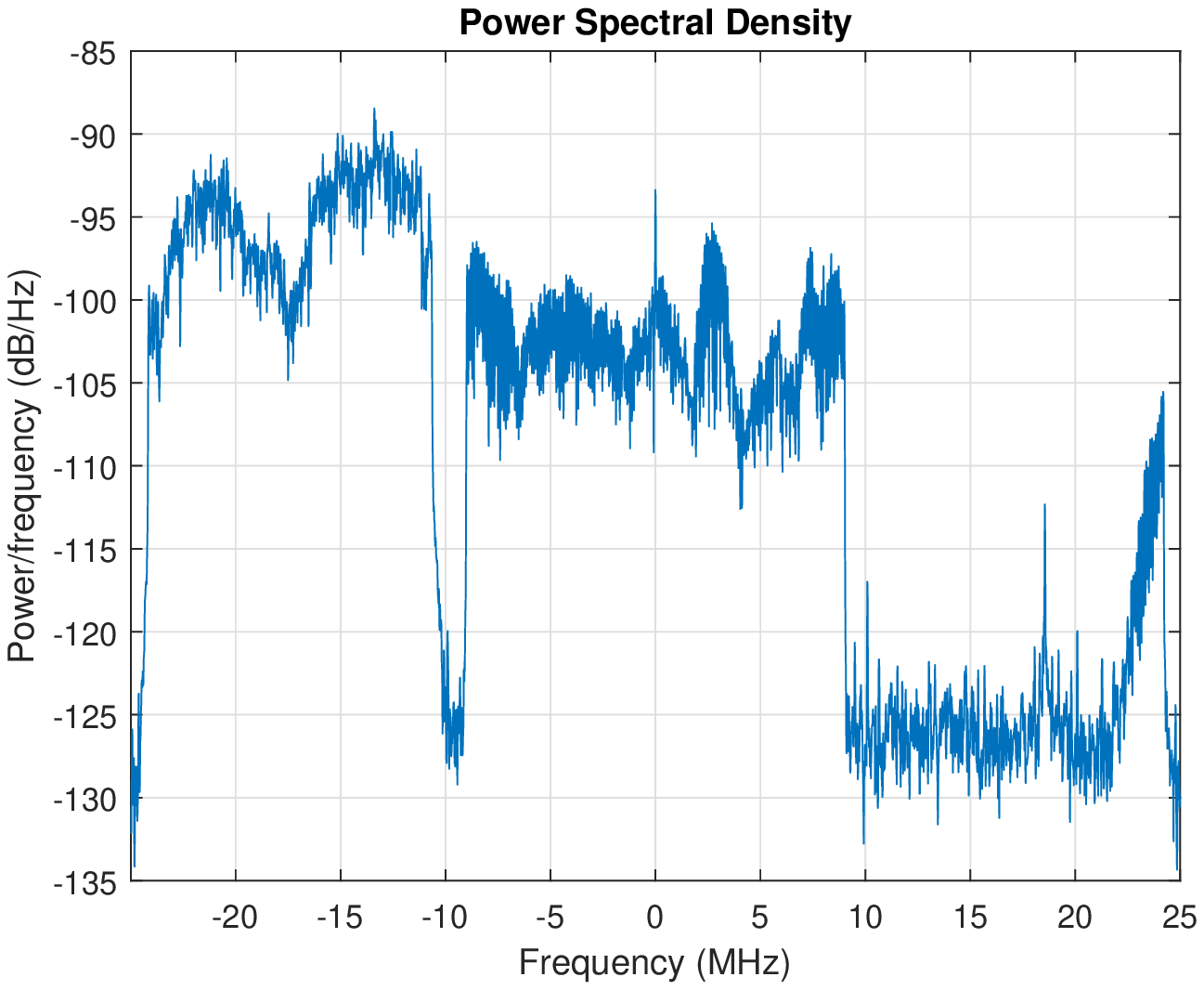}\label{fig:pwelch_lte}}\\
	\subfloat[UMTS]{\includegraphics[width=0.49\linewidth,clip=true]{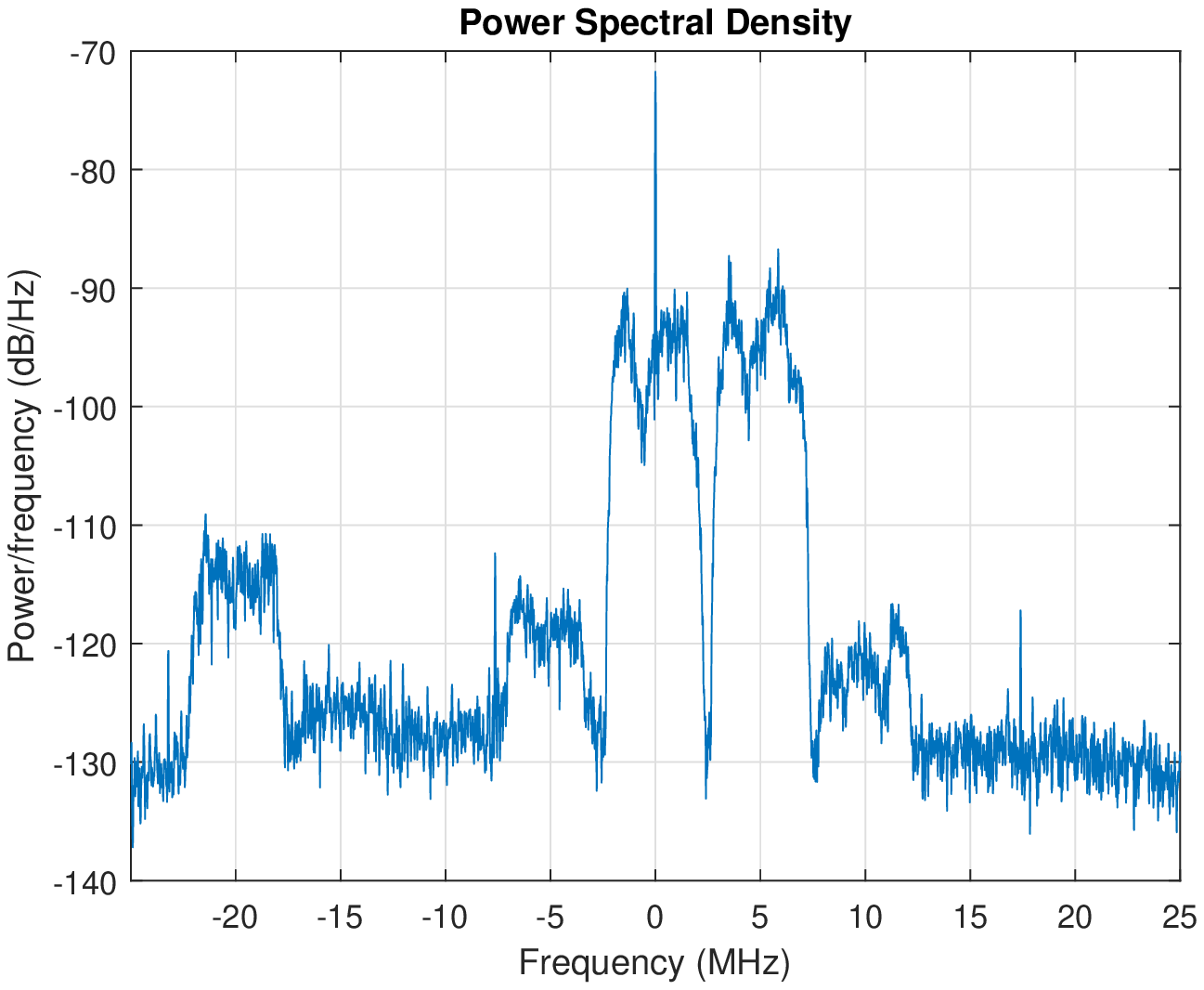}\label{fig:pwelch_umts}}
	\subfloat[GSM]{\includegraphics[width=0.49\linewidth,clip=true]{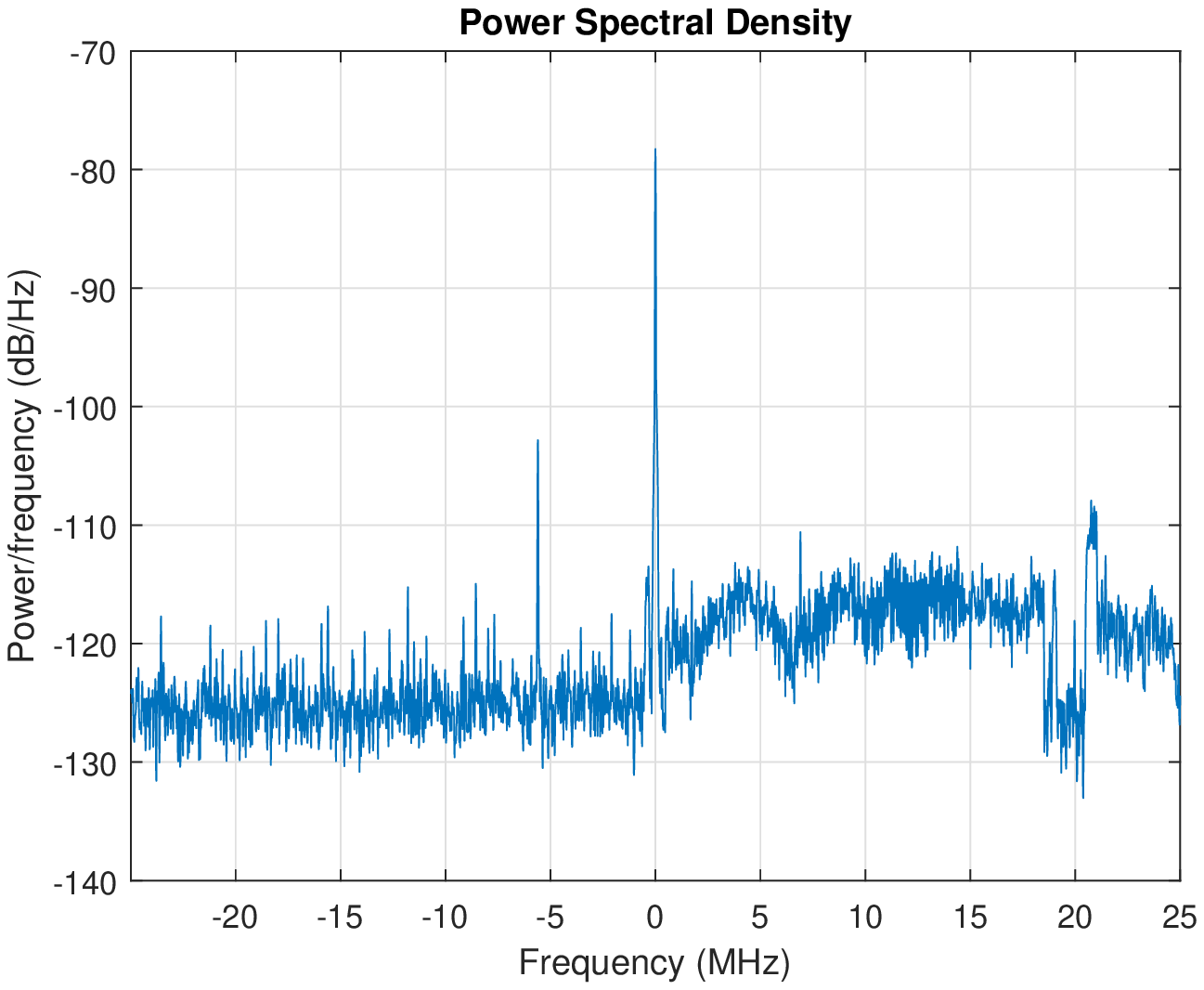}\label{fig:pwelch_gsm}}
	\caption{PSD of the employed sets of real IQ samples for each waveform.}
	\label{fig:pwelch_wv}
\end{figure}

Fig. \ref{fig:pwelch_wv} display the PSD of the captured signal corresponding to DVB-S2, LTE, UMTS and GSM bands, respectively. DVB-S2 is in the Ku band. LTE bands are in the 800 and 1800 MHz. UMTS bands are in the 700 and 2600 MHz. GSM bands are in the 700 and 1900 MHz.

As reported before, for this use-case we propose to use a LSTM ML model. The network is configured to use $3$ classes (LTE, UMTS and GSM) and $128$ hidden units. The output of the LTSM is connected to a fully connected layer, with $3$ classes. Finally, the output is connected to a soft-max layer in order to obtain the probabilities of interference classification. Based on these probabilities, the output is tagged with the type of interference.

The most critical part is the feature extraction, which is used as the input of the Neural Network. This task aims at pre-processing the data to obtain different metrics, which are used later by the LSTM. In our scenario, since we manipulate baseband digital radio signals, $4$ features are used:
\begin{itemize}
\item Magnitude of the temporal signal. 
\item Phase of the temporal signal.
\item Magnitude of the spectrum.
\item Phase of the spectrum.
\end{itemize}

\section{Results}
In this section, we describe the results that we obtained in both use cases: interference detection by using autoencoding and interference classification by using LSTM networks.

\subsection{Interference Detection}
As aforementioned, the MSE vector contains an error deviation between the input and output of autoencoder. Since it is trained to produce a signal with the same statistical properties, when the input is altered, the output is also altered. Thus, the statistical properties of the MSE vector are affected.

\begin{figure}[!ht]
	\centering
	\subfloat[PDF]{\includegraphics[width=0.49\linewidth,clip=true]{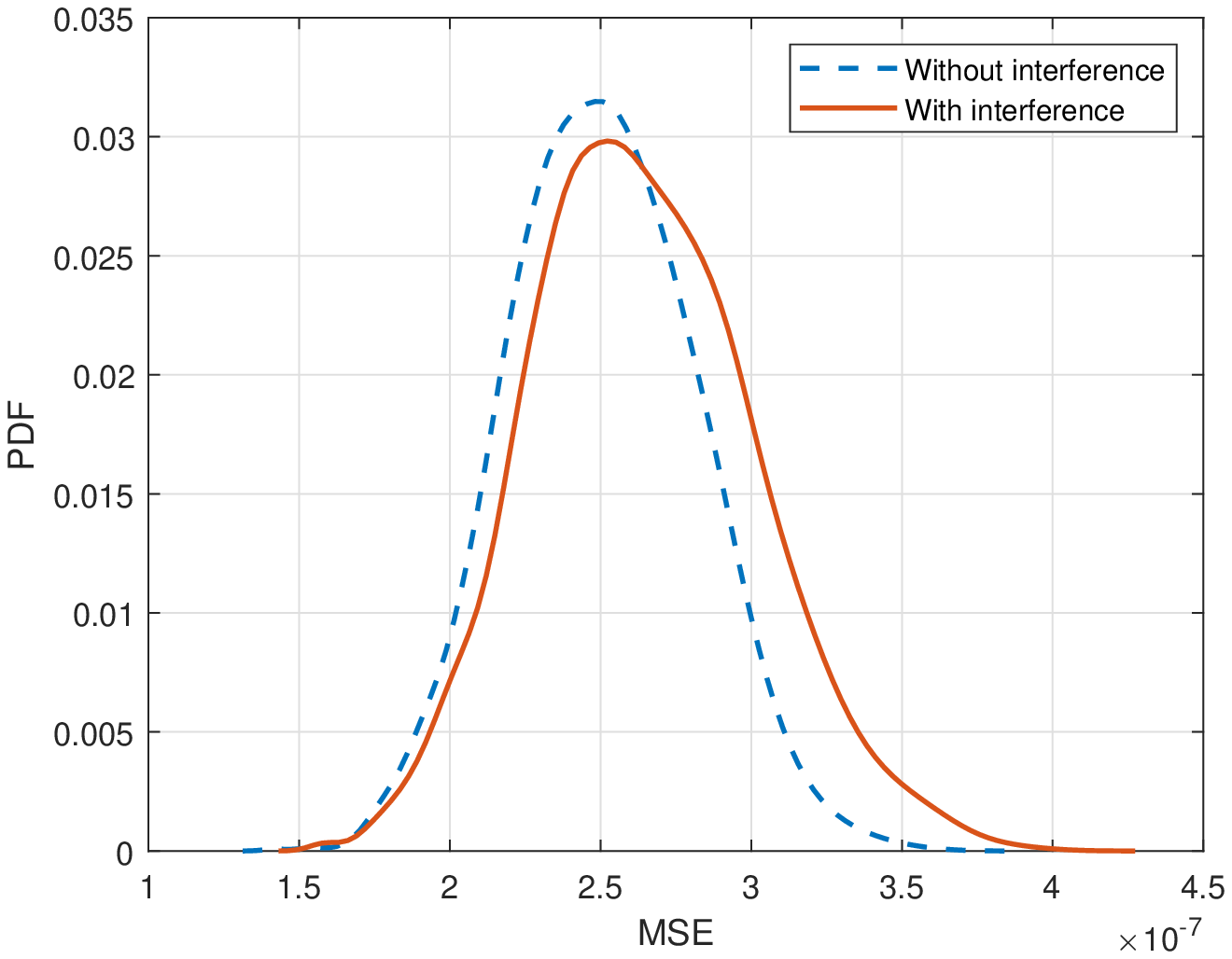}\label{fig:pdf_intf}}
	\subfloat[CDF]{\includegraphics[width=0.49\linewidth,clip=true]{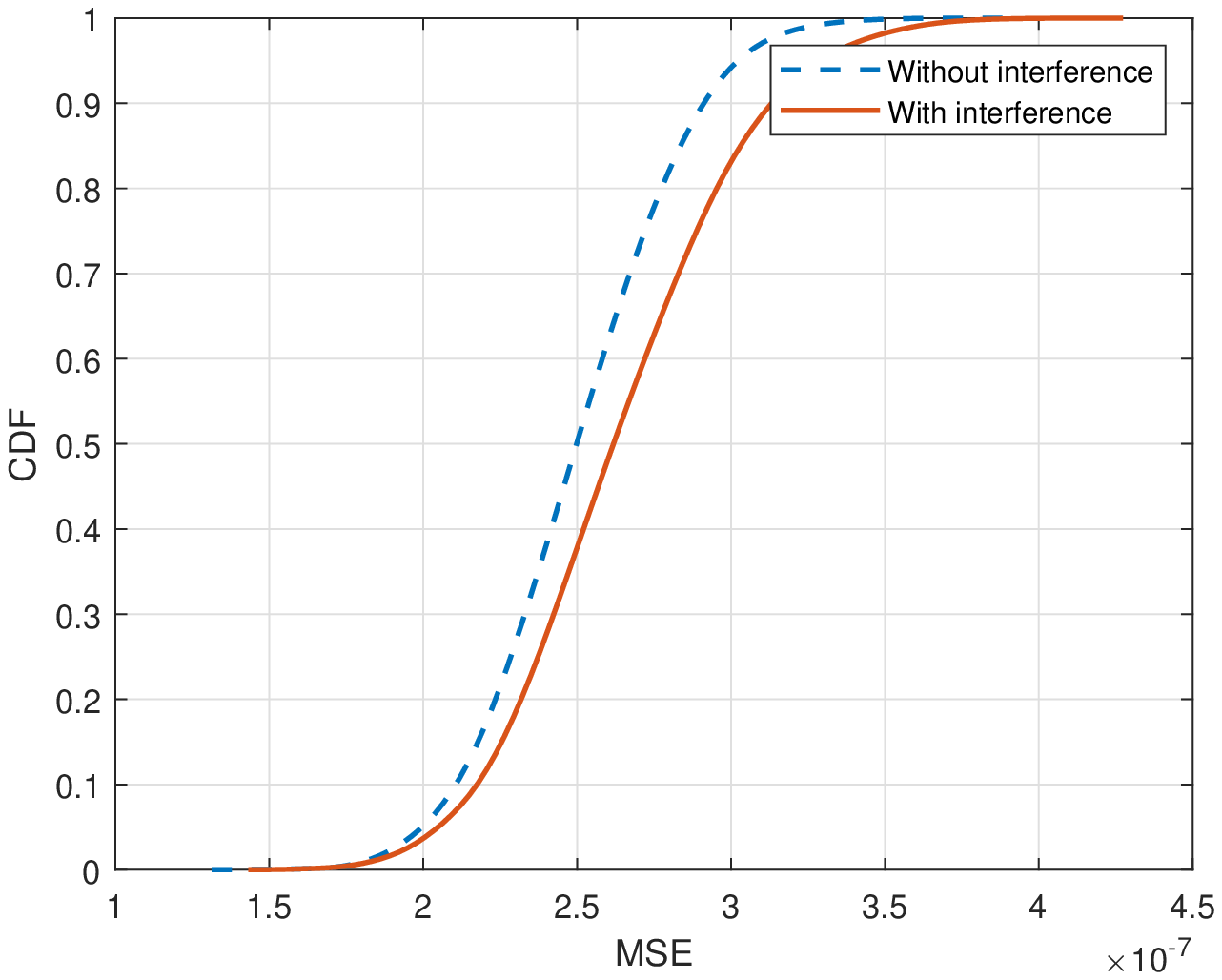}\label{fig:cdf_intf}}
	\caption{PDF and CDF of MSE vector.}
	\label{fig:pcdf_intf}
\end{figure}

However, visual inspections of the MSE vector or examinations of PSD (Fig. \ref{fig:pwelch_yesint}) are hard to use since are too vague and subjective. Instead, we use the statistical information of the MSE vector. Fig. \ref{fig:pcdf_intf} depicts the Probability Density Function (PDF) and Cummulative Density Function (CDF) of the MSE vector for signals with and without interferences.

It is clear that PDF and CDF of signals with and without interference are different and we can exploit these differences to define a detector. Aiming at producing an automatized process, we extract the statistical differences of MSE vector. Thus, we obtain the first four statistical moments of the MSE vector. Table \ref{tab:expr_4moms} describes the first four moments, where $\kappa_n$ is the $n$th sample of MSE vector and $N$ is the length of MSE vector.
\begin{table}[!ht]
	\centering
	\caption{Expressions of first four moments.}
	\begin{tabular}{|c|c|c|}
		\hline
		Moment & Name & Expression\\\hline
		\nth{1} & Mean & $\phi_1=\frac{1}{N}\sum_{n=1}^N\kappa_n$\\\hline
		\nth{2} & Variance & $\phi_2=\frac{1}{N-1}\sum_{n=1}^N\left(\kappa_n-\phi_1\right)^2$\\\hline
		\nth{3} & Skewness & $\phi_3\frac{1}{\sqrt{\phi_2^3}}\sum_{n=1}^N\left(\kappa_n-\phi_1\right)^3$\\\hline
		\nth{4} & Kurtosis & $\phi_4\frac{1}{\phi_2^2}\sum_{n=1}^N\left(\kappa_n-\phi_1\right)^4$\\\hline
	\end{tabular}
\label{tab:expr_4moms}
\end{table}

We compute the statistics of MSE vector using the expressions of Table \ref{tab:expr_4moms}. Results are summarized in Table \ref{tab:intdet_res} and unveil interesting remarks. The first moment order, the mean, does not show relevant aspects. In contrast to the mean, the variance and, specially, the skewness depict an important increase. When the input signal does not match with the trained, the error of the autoencoder becomes more random by increasing the spread of values (measured by the variance) and the asymmetry (measured by the skewness). Finally, the kurtosis does not have a relevant impact in the results. This is because the tailedness of the PDF does not vary.
\begin{table}[!ht]
	\centering
	\caption{Results of first four moments of MSE vector for signals with and without interferences.}
	\begin{tabular}{|c|c|c|c|}
		\hline
		Moment & No interference & With interference & Relative increase\\\hline
		Mean & $2.5\times10^{-7}$ & $2.6\times10^{-7}$ & $\uparrow\,+5.3\%$\\\hline 
		Variance & $9.4\times10^{-16}$ & $1.4\times10^{-15}$ & $\uparrow\,+44.3\%$\\\hline 
		Skewness & $7.5\times10^{-2}$ & $2.4\times10^{-1}$ & $\uparrow\,+219\%$\\\hline 
		Kurtosis & $2.9$ & $3$ & $\uparrow\,+3.1\%$\\\hline 
	\end{tabular}
	\label{tab:intdet_res}
\end{table}

To conclude this section, we aim at remarking that inspecting the error vector between the output and the input of the autoencoder is not sufficient to decide the presence of an interference. On the contrary, by inspecting the increase of the variance and the skewness of the error vector shows important differences that enables to detect and decide the presence of interferences.

\subsection{Interference Classification}
To train the LSTM network, we first segment the whole set of samples. The minimum unit is 512 samples, which are used to construct batches that are passed to the LSTM. The solver for the training is the Adam optimizer \cite{Kingma2014a}. In the next section, we deploy this technique to classify interference by using real captured signal.
\begin{figure}[!ht]
	\centering
	\includegraphics[width=0.9\linewidth,clip=true]{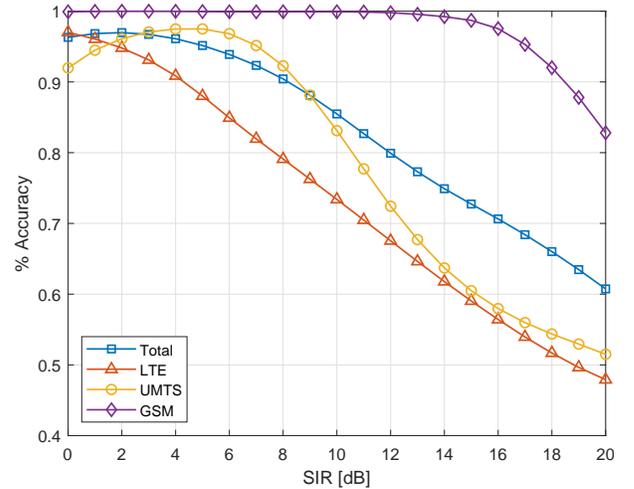}
	\caption{Accuracy of the interference classification.}
	\label{fig:intf_class_acc}
\end{figure}

Fig. \ref{fig:intf_class_acc} depicts the accuracy (true positives and true negatives) of the proposed framework. As expected, by the time that SIR increases, the performance decreases. For high SIR, the interference is weaker and, therefore, the classification accuracy decreases. In contrast, for low SIR, the interference is stronger and thus, the accuracy is near $100\%$.

In Fig. \ref{fig:intf_class_acc}, the individual accuracy of different terrestrial interferences is also depicted. Clearly, GSM obtains the highest accuracy compared with LTE and UMTS. UMTS accuracy is also important but drops more drastically for high SIR. Finally, LTE decreases linearly with SIR.
\begin{figure}[!ht]
	\centering
	\includegraphics[width=0.9\linewidth,clip=true]{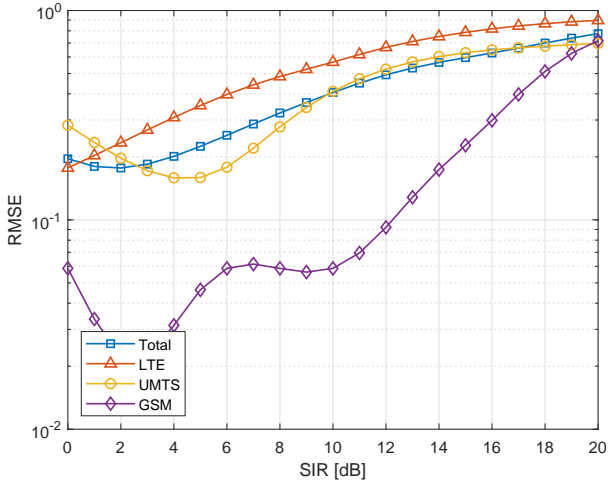}
	\caption{RMSE of the interference classification.}
	\label{fig:intf_class_rmse}
\end{figure}

Fig. \ref{fig:intf_class_rmse} illustrates the overall Root-Squared Mean Error of the proposed framework. As previously, the RMSE increases with SIR. Also, the individual RMSE is also depicted. Clearly, the GSM classification produces the less RMSE, followed by UMTS. LTE is the most difficult classification.

\begin{figure}[!ht]
	\subfloat[$\textrm{SIR}=0$ dB]{\includegraphics[width=0.49\linewidth,clip=true]{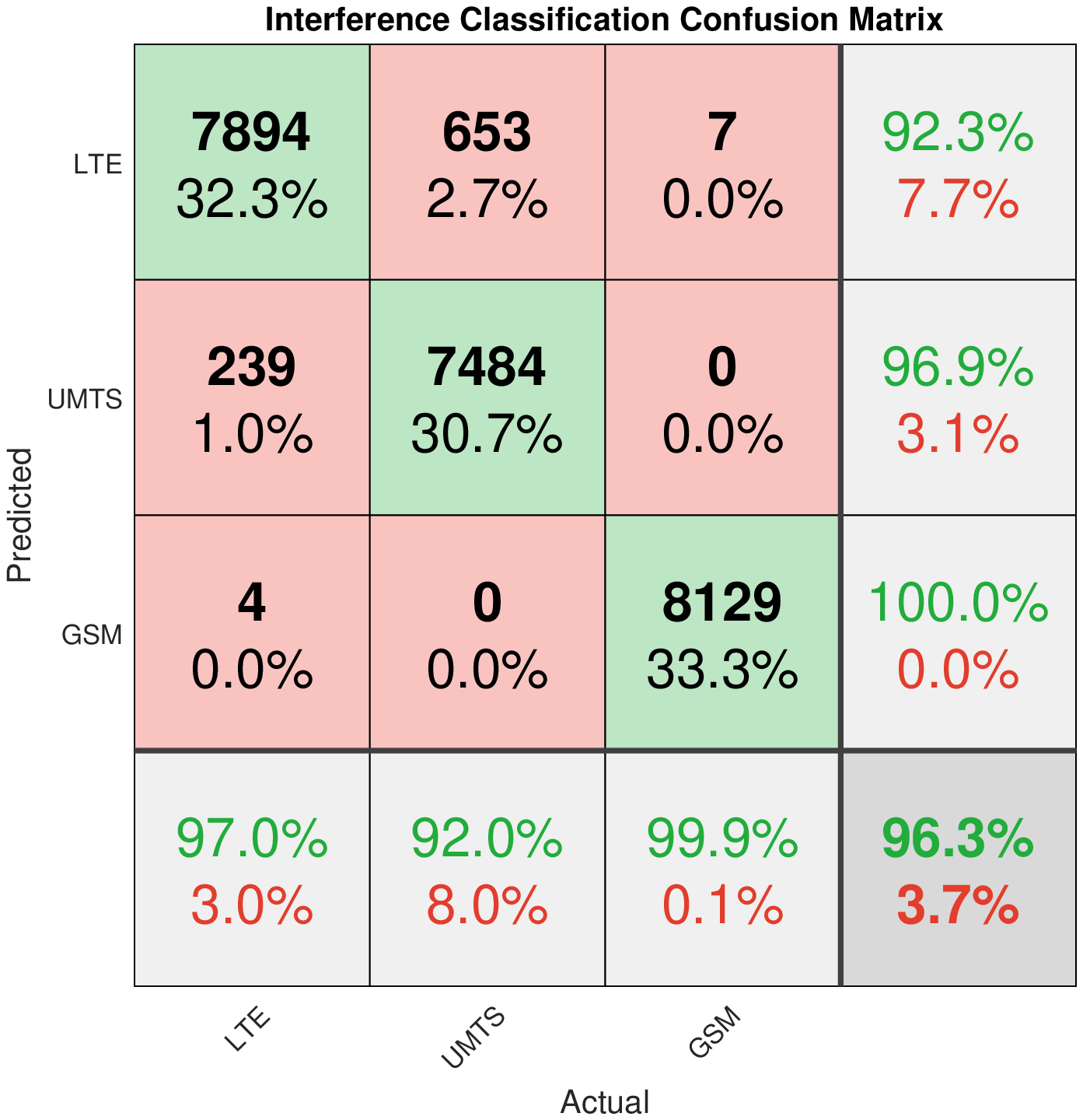}\label{fig:confmat_0}}
	\subfloat[$\textrm{SIR}=20$ dB]{\includegraphics[width=0.49\linewidth,clip=true]{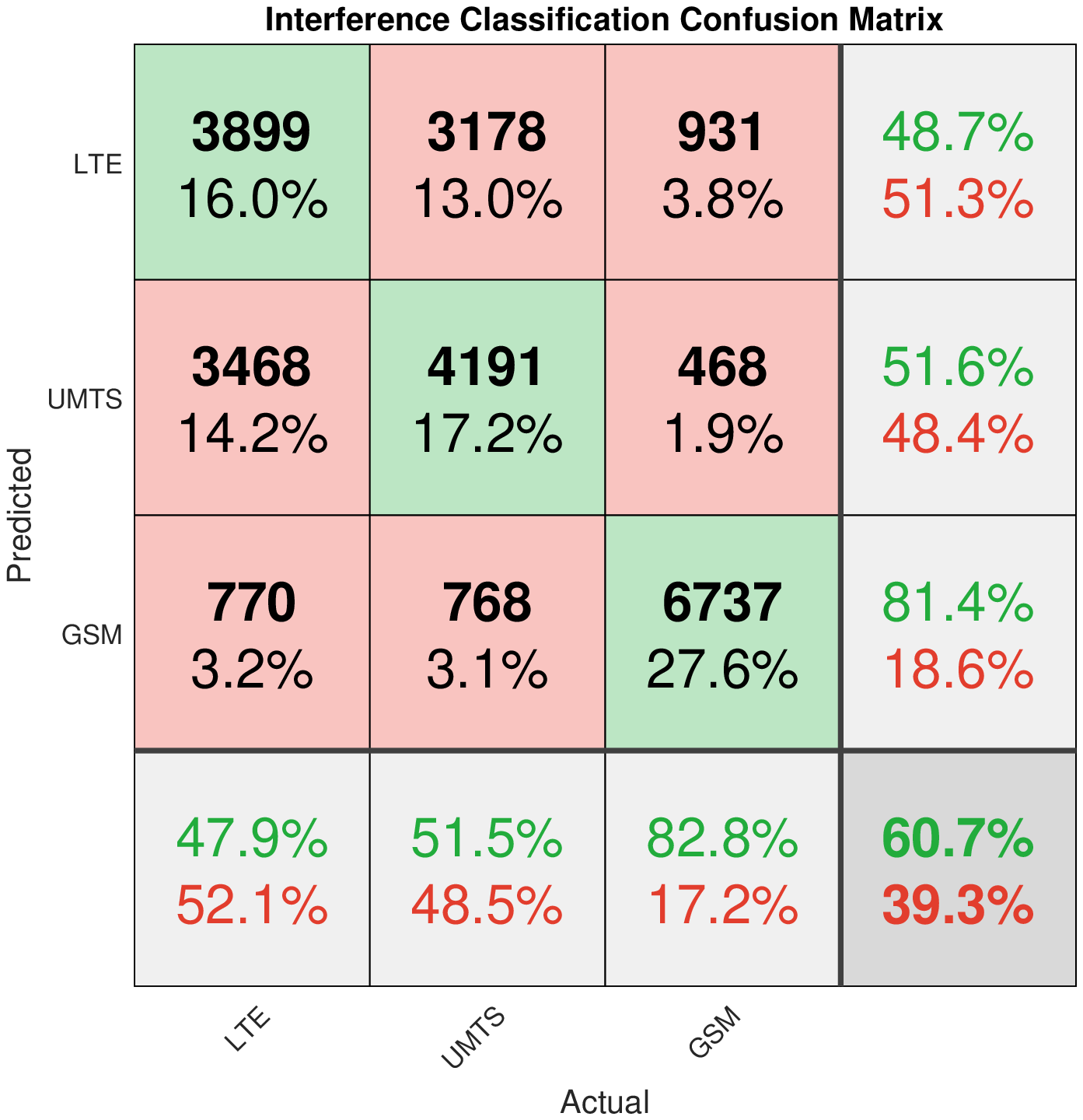}\label{fig:confmat_20}}
	\caption{Confusion matrix of the classificator.}
	\label{fig:confmat}
\end{figure}

The Confusion Matrix summarizes the performance between true positives, true negatives, false positives and false negatives, and also accuracy and error rates. This matrix provides information on how the classifier works and in which circumstances fails. It is also helpful for identifying possible biases in the prediction algorithm. Fig.\ref{fig:confmat} plots the confusion matrix of the classifier for SIR of $0$dB and $20$dB, respectively. As seen in previous figures, the classifier obtains more accurate results for low values of SIR and this can be examined in both confusion matrices.

\section{Conclusions}
In this paper we introduced Deep Learning algorithms to radio communications for interference management. It is based on two major areas: interference detection and interference classification. The former uses autoencoding techniques in order to decide whether interference is present or not. The latter is employed when the interference detector decides and it classifies the interference among predefined standards. The classificator can be trained with different standards in order to cope with as many types of interferences are considered. The results show relevant aspects and are promising to deal with the forecoming communications. In particular, the detector is able to detect interferences regardless their bandwidth or frequency position. The classificator is able to estimate which type of interference is. Finally, we illustrate the performance of both areas by different levels of the power of the interference. 

\printbibliography

\end{document}